\documentclass[preprint,12pt]{elsarticle}
\usepackage[T1]{fontenc}
\usepackage{amssymb}
\usepackage{graphicx}
\journal{Physica A}
\begin{document}

\begin{frontmatter}

\title{Probability of Large Movements in Financial Markets}
\author[ioc]{Robert Kitt\corref{cor1}}
\ead{kitt@ioc.ee}
\author[ioc]{Maksim S\"akki}
\author[ioc]{Jaan Kalda}
\address[ioc]{Department of Mechanics and Applied Mathematics, Institute of Cybernetics at Tallinn University of Technology, 12061, Tallinn, ESTONIA}
\cortext[cor1]{Corresponding author. Tel.: +372 6204174; fax: +372 6204151}

\begin{abstract}
Based on empirical financial time-series, we show that the "silence-breaking" probability 
follows a super-universal power law: the probability of observing a large movement is {\em inversely proportional to the length of the 
ongoing low-variability period}.
Such a scaling law has been previously predicted
theoretically \cite{kitt2005}, assuming that the length-distribution of the low-variability periods follows a 
multiscaling power law. 

\end{abstract}

\end{frontmatter}

\section{Introduction}
\label{intro}

The power laws and scaling behaviour are present in numerous aspects of human societies 
and in the nature. One of the first promoters of the concept of the power law was Vilfredo Pareto (cf.~\cite{pareto}), 
who studied the wealth distribution in different societies. Further, Harvard linguistics professor George Zipf, again a representative of social sciences, 
observed that only few words in English language are used very often, and most of the words are used rarely. (cf.~\cite{zipf1949}). 
Nowadays, physicists  are very used to the power-laws, which,  however, are sometimes somewhat counter-intuitive.
Indeed, the presence of a power-law means that there are some representatives of a population, which  are very different from the typical members of that population.
For example, as  the result  of a social evolution, the wealth of a single individual can qualitatively change the wealth of a large community.
This is completely different from the biological evolution: e.g. the weight of a single living creature makes always only a tiny contribution to the net biomass of the corresponding (non-small) population.

The presence of a wide spectrum of power laws in finances can be ascribed to the fact that socially interacting humans form large complex systems, which 
are characterized by self-organized criticality \cite{Bak}.
This makes finances (alongside with the turbulence) a fruitful polygon for studying various aspects of scale-invariance.
Indeed, various power-laws have been observed in financial time series since 1960-es, by B. Mandelbrot (cf.~\cite{mandelFGN,mandel97} and references therein). In 1999, Ausloos and Ivanova also reported the multifractality in financial time series (cf.~\cite{ausloos}). 
Around 1990-ies, the studies of scale-invariance in finances became more extensive, c.f.\  \cite{stanley,voit,takayasu,bouchaud}, effectively creating a new branch of 
statistical physics --- the \emph{econophysics}. A recent overview of the progress in understanding 
the scaling and its universality in finances can be found in Ref.\ \cite{stanley2008}. The aim of the current study is 
to contribute to the {\em understanding of the origins of universality}. 
Mathematically, our basic idea is very simple, nearly trivial. However, when dealing with the sources of universality, mathematically simple and robust models have better chances of describing reality, than complex and elaborate constructions; c.f. the Occam's razor.

The attempt to successfully and systematically predict the direction of future movements of asset prices can be compared 
to the attempts of inventing the \emph{perpetum mobile}. However, the attempt to characterize and predict the risk (or volatility) 
may offer significantly better results and therefore the volatility is one of the most-studied phenomena in Econophysics. 

\begin{figure}[htp]
	\centering
	\includegraphics[width=1.0\textwidth]{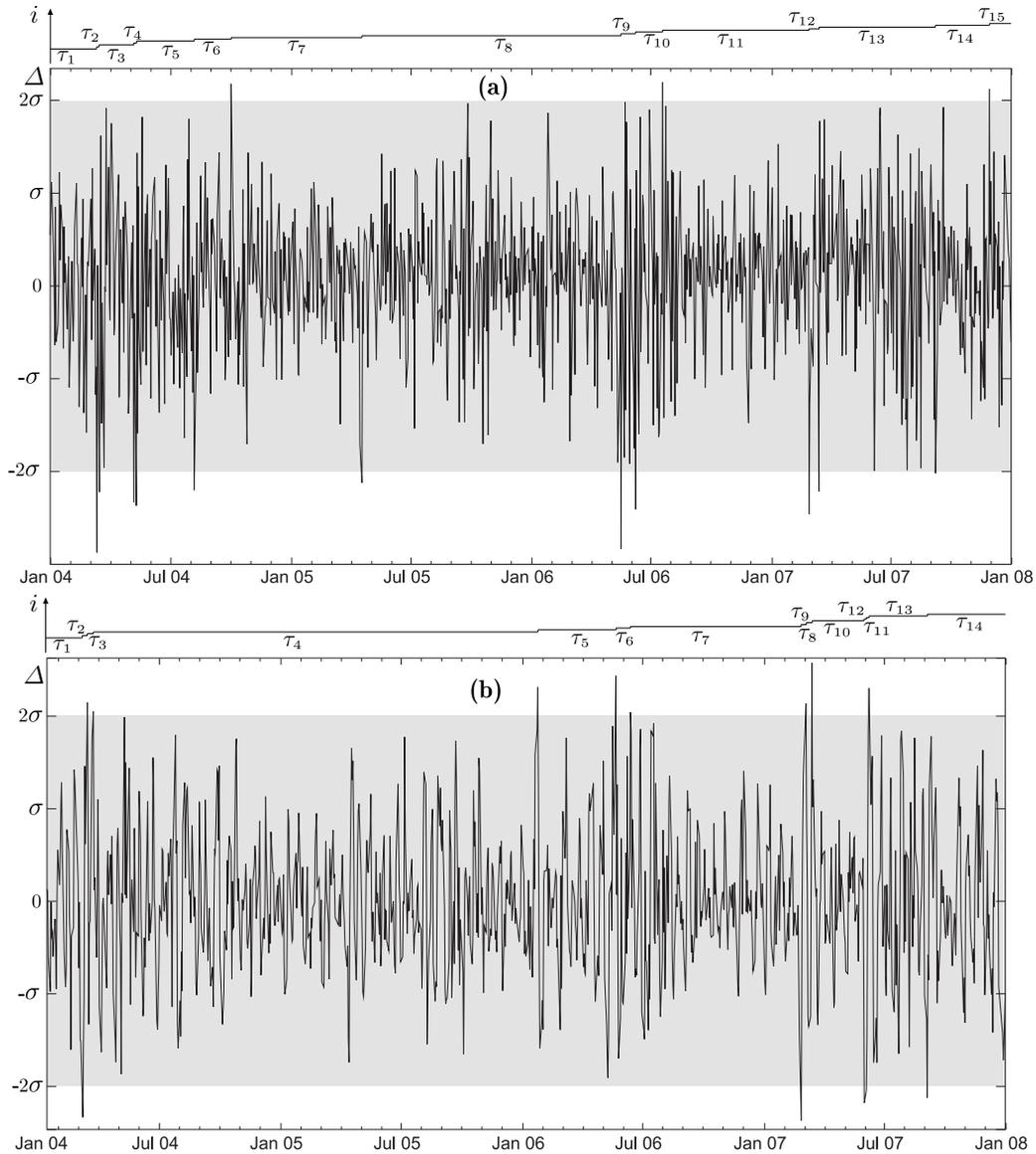}
	\caption{Variability $\Delta$ of the DAX index and the respective low-variability periods for 4 years (2003-2007), 
	using $w=1\,$ day (a) and $w=10\,$days (b).
	For a pre-fixed threshold level $\delta= 2\sigma$ (where $\sigma$ is the standard deviation of the signal), 
	low-variability intervals of duration $\tau_i$ are formed 
	as the  intervals corresponding to such graph segments, which lay 
	entirely inside the gray area. The small graphs in top share the time axis with the bottom graph, and 
	illustrate the fragmentation of them into low-variability intervals by plotting the interval index $i$ versus time $t$.}\label{fig:tau}
\end{figure}

One of the most challenging features of the volatility dynamics are the intermittently appearing extreme price movements, which are often 
accompanied by overall increase of volatility over a certain time window. Traditionally, such a behaviour has been described by multifractal spectra.
The multifractal analysis is untoubtedly a powerful tool; however, due to the involved mathematical methods, it is not well-suited for practical applications of the prediction and optimisation of risks. This observation motivated us to introduce a complementary method of the length-distribution analysis of the {\em low-variability periods} \cite{kalda99,sakki2004,kitt2004}. The low-variability periods are defined as consequent time periods, during which
the price changes of the observed asset (as compared to the local average over a sliding window of width $w$) remains under a pre-set threshold $\delta$. The illustration of the method is shown in Fig.~\ref{fig:tau}.
This method (with certain modifications) was developed independently within different contexts, and put under extensive tests by several research groups.
So, the effects of the long-term memory and clustering of extreme events in various time series (cf.~\cite{bunde2005,livina2005,bogachev2007,bogachev2008,eichner2007}), are, in fact, closely related to  (and in a certain sense covered by) the low-variability period analysis.
The same applies to the studies of the time intervals $\tau$ between \emph{volatilities} which are above a threshold $q$
\cite{wang2006,wang2007,weber2007,wang2008,vodenska2008,jung2008}).

\noindent

Our previous studies \cite{kitt2004} and \cite{kitt2005} have shown that 
\begin{itemize}
\item financial time-series are typically characterized by multi-scaling behaviour of the low-variability periods (power law can be observed for 
a certain range of the parameters  $\delta$ and $w$, and the scaling exponent depends on these  
parameters);
\item theoretically, a multi-fractal time-series follows also  a multi-scaling behaviour of the low-variability periods; the 
scaling exponent can be expressed via the multifractal exponents;
\item as compared to the multifractal analysis, the analysis of the low-variability periods is easier to implement, has higher resolution of time-scales, 
and the results of the analysis can be interpreted more straightforwardly.
\end{itemize}
These findings agree well with the above cited independent studies.

The market fluctuations have been also modelled as L\' evy flights and continuum time random walks (CTRW); 
therefore it is also of interest to mention that in the case of uncorrelated L\' evy flights, 
there is no power-law for the length-distribution of the low-variability periods:
length-distribution decays exponentially; the same applies to Gaussian time-series, e.g. (non)persistent random walks.
Meanwhile, in the case of CTRW, there is 
a mono-scaling behaviour of the low-variability periods: the scaling exponent does not depend on the parameters  $\delta$ and $w$, and is defined
by the exponent of the waiting-time distribution \cite{kitt_up}.  

In addition to the above listed results, we have shown \cite{kitt2005} that the very presence of a power-law for the probability distribution function of the low-variability segment lengths (even if observed for a narrow range of the parameters $\delta$  and $w$) bears interesting consequence: 
the probability $p$ of observing a large movement (exceeding the threshold parameter $\delta$) in the time series during the next period  of duration $w$, is inversely proportional to {\em the length of the on-going low-variability period} 
(i.e. to the time elapsed since  the most recent large movement; note that by definition, $p<1$, and $p=1$ would imply that for the given parameters, the next price movement exceeds definitely the threshold $\delta$). This super-universal scaling law has been derived independently 
by Bogachev, {\it et al} \cite{bogachev2007,bogachev2008}.
Here we repeat the derivation briefly.

First, we assume that the length of the periods is measured in the units of the window length $w$.
Then, the probability of a large movement during the unit time is given by the ratio of {\em (a)} the number of those low-variability periods, 
the length of which is exactly $n$, $N_a=R(n)-R(n+1)$, and {\em (b)} the number of those low-variability 
periods, which have length $m\ge n$, $N_b=R(n)$.
Here, $R(n)$ denotes the cumulative length-distribution function of the low-variability periods with length equal to $n$ 
[i.e.\ $R(n)$ represents the number of low-variability periods with length greater or equal to $n$ time units]. So, we can 
express the probability of "silence-breaking" (i.e. the probability that the next movement exceeding the threshold $\delta$) as a function of the on-going "silent" period $n$:
\begin{equation}
p(n)=[R(n)-R(n+1)]/R(n).
\end {equation}
If $n$ is large, the difference $R(n)-R(n+1)$ can be calculated approximately as $-\frac {dR}{dn}$. 
Upon applying the power-law $R(n)=R_0n^{-\alpha(\delta,w)}$, we arrive at $N_a\approx \alpha R_0 n^{-\alpha-1}$ and $N_b\approx R_0 n^{-\alpha}$. 
Bearing in mind that $p(n)=N_a/N_b$, the final result is written as 
\begin{equation}\label{silence}
p(n) \approx \alpha n^{-1}.
\end {equation}
Since the scaling exponent of this law is independent of the parameters $\delta$ and $w$ (as well as of the scaling exponent $\alpha$), 
it can be called super-universal.
The equation (\ref{silence}) can be also denoted as \emph{the law of the silence-breaking probability}, because it describes the probability 
of a large movement after a longer ``silent'' period (characterized by small movements). 
Recently, similar application of the risk measurement was proposed by Wang, {\it et al} \cite{wang2007}.

As demonstrated above, the mathematics behind this super-universal law is extremely simple. However, 
we do believe that the consequences of it are profound, and it allows to shed light into the 
origin of universality in the dynamics of market fluctuations. In fact, 
a very similar behaviour (with a similar mathematical origin) has been detected in the context of 
direct avalanches in self-organised critical systems (c.f.~\cite{maslov}), and proven to be a useful tool in 
understanding the universality in burst dynamics. These arguments motivated us to put equation (\ref{silence})
under several tests, using both the data of real market fluctuations, as well as the surrogate data. 
{\em The aim is to clarify, how well is the 
super-universal law followed by the real-world time-series}, for which the scaling laws are far from being perfect (and is for certain applications better described, e.g., by stretched exponentials \cite{bunde2005}), and
for which the finite-size effects can (possibly) mask the theoretical scaling behaviour.
In order to show the presence of the $1/n$-law in relatively short time-series 
(with unavoidably poor statistics of long low-variability periods), 
an appropriate data analysis technique has been developed (see Section \ref{constraints})

\section{Empirical tests}
\label{empirics}
\begin{figure}[htp]
	\centering
	\includegraphics[width=1.0\textwidth]{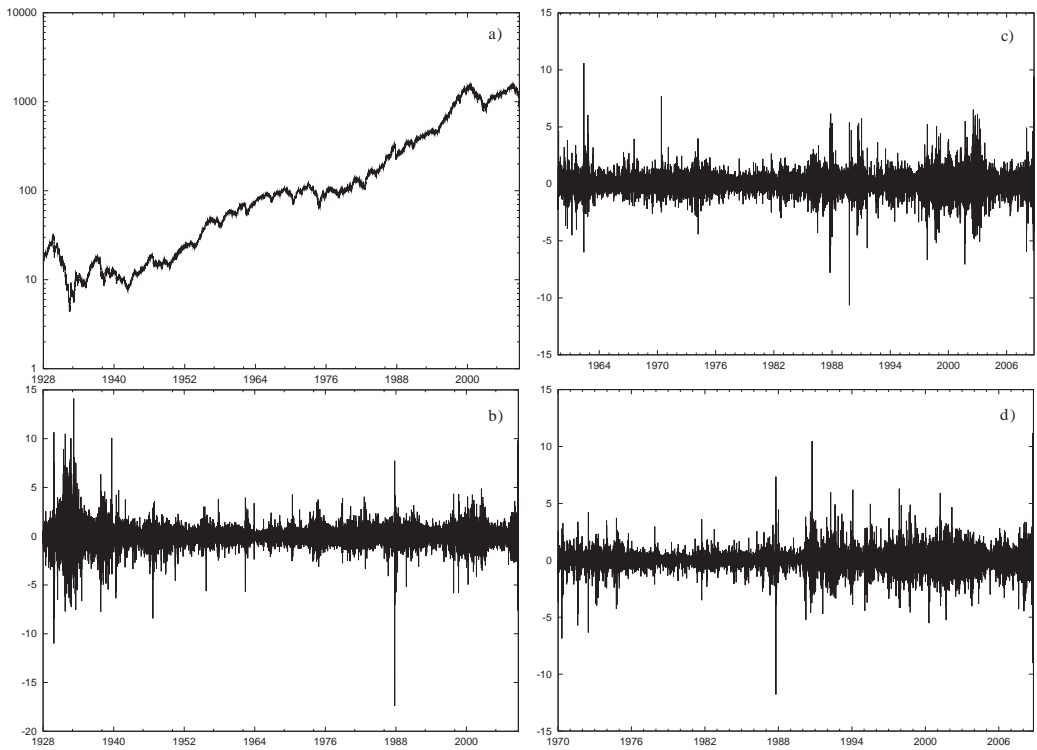}
	\caption{a) S\&P~500 index; daily returns [measured in standard deviations] of S\&P~500 (b), German Stock Index DAX (c), and Nikkei 225 Index (d)}\label{fig:spx-signal}
\end{figure}

Further we proceed to test empirically the probability of the "silence-breaking". 
As changes in financial markets are believed to obey power-laws (\cite{gabaix}), we use several financial time series (cf. Table \ref{data}) 
in order to test the equation (\ref{silence}). Some of these data are presented visually in Fig.~\ref{fig:spx-signal}.
We also test the equation (\ref{silence}) with artificially generated time series (i.e. surrogate data with known properties, see Appendix), 
which obey a power-law.

\noindent

\begin{table}[loc=htbp]
\scriptsize
\caption{The data used in empirical analysis (source of all the data series is Bloomberg)}
\label{data}
\begin{tabular}{|l|p{5.5cm}|c|r|} 

\hline
Abbr&Description&Calendar Period&\# of trading days\\ \hline
SPX&Standard \& Poor's 500 Index&30/12/27 - 31/10/08& 20305\\ \hline
DAX&The German Stock Index&01/10/59 - 31/10/08& 12326\\ \hline
NKY&Nikkei 225 Stock Average&05/01/70 - 31/10/08& 9588\\ \hline
MXEA&The MSCI Europe, Australasia and Far East Index&31/12/71 - 31/10/08& 9610\\ \hline
CAC&CAC-40 Index of Paris Bourse&09/07/87 - 31/10/08& 5374\\ \hline
UKX&FTSE 100 Index&03/01/84 - 31/10/08& 6282\\ \hline
MXWO&MSCI World Index&03/01/72 - 31/10/08& 9610\\ \hline
INDU&Dow Jones Industrial \-Average&03/01/1900 - 31/10/2008& 27333\\ \hline

\end{tabular}
\end{table}

\begin{figure}[htp]
\centering
\includegraphics[width=0.7\textwidth,angle=270]{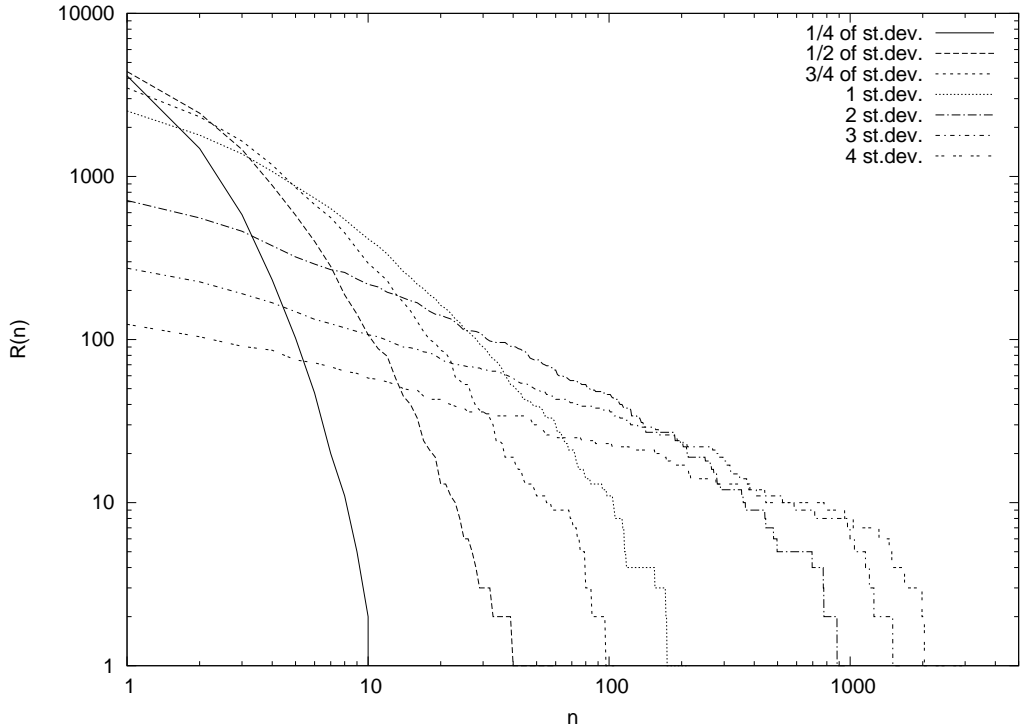}
\caption{Cumulative distributions of low-variability periods $R(m \ge n)$ over the length $n$ for the S\&P~500 time series, using different threshold levels}\label{fig:lvp-distr-all-thresholds}
\end{figure}

We analysed the daily returns of eight different indices of the world stock exchanges (cf. Table \ref{data}). 
Cumulative distributions $R(n)$ for the S\&P~500 index for various thresholds $\delta$ are shown in Fig.~\ref{fig:lvp-distr-all-thresholds}; the other indices follow a similar behaviour.
The values of $\delta$ were normalized to the standard deviation, which was calculated for the whole sample of every index.
We found that $R(n)$ was in a good accordance with the power-law for a broad range of thresholds $\delta$ (ranging from 3/4 to 4 standard deviations).
For smaller thresholds, inertial range of the linear part in the log-log plot was too short for a meaningful scaling analysis.
On the other hand, greater values of $\delta$
produced relatively small total number of low-variability intervals, insufficient for the further analysis (cf. Table \ref{delta-choice}). 
Therefore, the optimal range for the threshold level was found to be between 1 and 2 standard deviations ($\sigma$). 
Here we have adopted the value $\delta= 2\sigma$.

\noindent

While the details of the functional relationship $\alpha(\delta, w)$ depend on the specific features  of the (non-Gaussian) statistics of the returns, 
generic trends can be outlined as follows.
Within the optimal range of the parameters, larger values of the threshold $\delta$ lead to longer low-variability periods, and to
a decrease of the overall number of the low-variability periods.
The value of $\alpha$ is the (modulus of the) slope of the straight part of the $R(n)$ curve in log-log graph (Fig.~\ref{fig:lvp-distr-all-thresholds}), which typically 
decreases for such a process (i.e. with increasing $\delta$). 
Similar arguments lead to the conclusion that larger values of the sliding window width $w$ lead to shorter low-variability intervals 
and hence --- to larger values of $\alpha$.

\begin{table}[loc=htbp]
\centering
\scriptsize
\caption{Some characteristics of the distribution function $R(n)$ for the  S\&P~500 index, using different values of the threshold parameter $\delta$ and $w=1\,$day. $N$ denotes the length of the longest low-variability period.}
\label{delta-choice}
\begin{tabular}{|r|r|r|r|}
\hline
$\delta$, in stddev&$R(1)$&$R(10)$&$N$\\ \hline
1/4&4145&2&10 \\ \hline
1/2&4394&107&59 \\ \hline
3/4&3491&294&105 \\ \hline
1&2523&414&215 \\ \hline
2&713&218&949 \\ \hline
3&274&107&1609 \\ \hline
4&124&58&3086 \\ \hline
5&67&32&6382 \\ \hline
\end{tabular}
\end{table}

The empirical tests of Eq.~(\ref{silence}) were conducted as follows: 
$(i)$ we count the number of low-variability periods $R(n)$ with length equal to $n$; $(ii)$ we calculate the probability $p(n)=[R(n)-R(n+1)]/R(n)$ 
and $(iii)$ plot it against $n$. In order to make the data presentation visually easier to understand, we modify step $(iii)$ by replacing $p(n)$ with $p(n) \times n$. 
Thus, should the equation (\ref{silence}) hold, the plot of $p(n) \times n$ against $n$ should yield a horizontal line.

\subsection{Finite-size effects}
\label{constraints}
Before proceeding to the results of the tests outlined above, it should be noted that the probability of silence-breaking 
cannot be expected to hold for all the values of $n$. We explain this claim in more details.

\begin{enumerate}
\item 	
Let us recall that in order to derive the scaling law (2), we approximated $R(n)-R(n+1)$ with $-\frac {dR}{dn}$;  this is, however, legitimate only for large values of $n$. 
Therefore, one can expect that the super-universal power law (2) fails for small values of $n$.
The surrogate data analysis confirms that this is, indeed, the case, see Fig.~\ref{fig:surrogate-alphas}.
Using numerically generated very long data-series [which correspond to the  distributions $R(n)=R_0n^{-\alpha}$], we have calculated the quantity  
$p(n) \times n$, and plotted it against $n$; in a full agreement with the theoretical expectations, it reaches asymptotically (for $n \gg 1$) a plateau with a limit value $p(n) \times n \to \alpha$.

\begin{figure}[htb!]
\centering
\includegraphics[width=0.7\textwidth,angle=270]{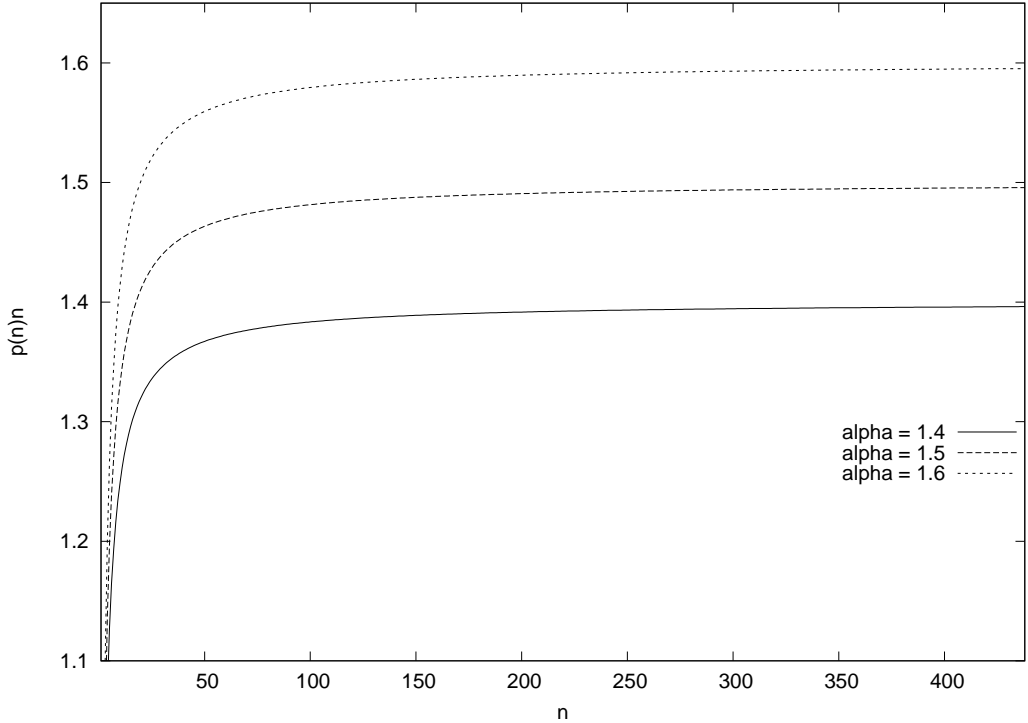}
\caption{The quantity $p(n) \times n$ is calculated for very long surrogate data series, and plotted against the length $n$; the results confirm the analytically 
predicted asymptotics $p(n) \times n = \alpha$ for $n \gg 1$}\label{fig:surrogate-alphas}
\end{figure}

\item The analysis of the empirical data shows that $R(n)$ is not a smoothly decreasing function of $n$. Instead, for larger values of $n$, there are long plateaus with a constant value of $R(n)$.
Indeed, it is possible that for many values of $n$, there is no low-variability periods with the given length $n$. 
It is apparent that due to the  finiteness of the  data series, the higher values  of $n$ offer less statistics, 
i.e. the likelihood of observing a low-variability period of a length, equal to a fixed and very high value of $n$, is very low.
Consequently, if we apply Eq.~(1) directly to the empirical data series, the resulting empirical probability estimates  $\tilde p(n)$ follow a very singular behaviour: there are sharp peaks at those values of $n$, 
which happen to have a matching low-variability period; between these peaks, $\tilde p(n)\equiv 0$. Evidently, such a simple-minded data analysis is not suited for testing the scaling law (2).

Perhaps the  most straightforward solution here would be to average the empirical probability estimates over a sliding window of a suitable length. However, this is certainly not the easiest and most elegant route, because 
it raises several technical issues (e.g. what is the best window width; it should variable and increase with $n$). Here, we have proceeded as follows. First, we approximate the step-wise experimental 
distribution law $R(n)$ with a piece-wise linear function $\tilde R(n)$. More specifically, for all these values of $n$ ($n=n_1, n_2,\ldots n_k$) for which $R(n+1)\ne R(n)$, 
we define $\tilde R(n)=R(n)$; for all the other values of $n$ we use a linear interpolation: for  $n\in (n_i, n_{i+1})$, we define 
\begin{equation}\label{piecewise-lin}
\tilde R(n)=R(n_i)(n_{i+1}-n)+ R(n_{i+1})(n-n_{i})/(n_{i+1}-n_i).
\end{equation}
Then, we can apply Eq.~(1) to the smoothed distribution function $\tilde R(n)$. 

\item Finally, even if we apply the smoothing procedure as described above, for very large values of $n$ (of the order of the length $N$ of the longest low-variability period), 
the statistical basis becomes so narrow that the function $p(n)=[\tilde R(n)-\tilde R(n+1)]/\tilde R(n)$ will fluctuate 
with a very large amplitude. Apparently,  the statistical fluctuations around the theoretical
power-law [Eq.~(\ref{silence})] grow with $n$; these fluctuations set an effective upper limit for the 
empirical scaling range at $n_{max}\approx N$.
\end{enumerate}

\begin{figure}[htb!]
\centering
\includegraphics[width=0.7\textwidth,angle=270]{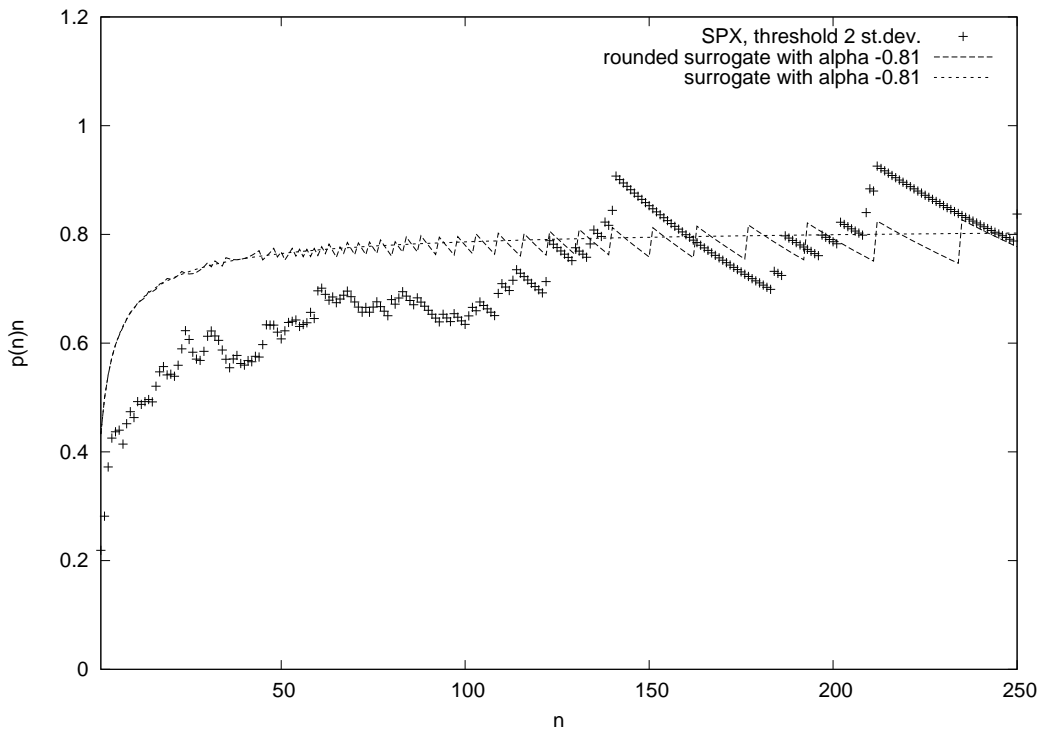}
\caption{The quantity $p(n) \times n$ is plotted for the S\&P~500 index ($\delta = 2\sigma$, $\alpha =0.81$) and for the numerically generated time series (of a similar length and with the same value of $\alpha$) against the period length $n$. 
These results are in a good agreement with the theoretical expectation: for $n\gg 1$, the graphs fluctuate around the asymptotical value $p(n) \times n\to \alpha$.}\label{fig:p-n-SPX-2stddev}
\end{figure}

Being guided by these considerations, we calculated the $R(n)$-functions for all the empirical data series, 
and applied the linearization procedure according to Eq.~(\ref{piecewise-lin}).
The results for the S\&P~500 are presented in Fig.~\ref{fig:p-n-SPX-2stddev}; the graphs for our other empirical data series behave similarly.
Note that in Fig.~\ref{fig:p-n-SPX-2stddev}, we have limited the data to $n\le N/4$, due to the reasons discussed in the previous pragraph.
As predicted theoretically, the grahps fluctuate around a  constant value [for the parameters of Fig.~\ref{fig:p-n-SPX-2stddev}, $p(n)n\approx 0.8$].
\noindent


\section{Discussions: the consequences of the power-law}
\label {discussions}

Using the financial time-series, as well as artificially generated (surrogate) data series, we have shown that   the "silence-breaking" probability 
follows, indeed, a super-universal power law: the probability of observing a large movement is {\em inversely proportional to the length of the 
ongoing low-variability period}. The usefulness of this result is not limited to the financial risk analysis, and helps to understand the origins of universality in the scaling behaviour of complex systems;
it can be also used to build a hierarchical risk prediction scheme (for a spectrum of events of different amplitude).

On the one hand, our result ``approves'' the common human tendency to forget those extreme events, which have happened a long time ago.
Indeed, we have shown that in any system producing a  scale-invariant sequence of events of different amplitude, 
the probability of observing a large event on next day is inversely proportional to the time elapsed since the last large event. 
On the other hand, however, our finding does not void another universal property of power laws: sooner or later, there will be an event of an even larger amplitude.

So, in the case of  stock markets, 
we can conclude that if we have not seen a  ten-percent-or-larger movement for the last 10~000 days, the probability of such a movement on next day can be estimated as $10^{-4}$. 
However, one should not be driven to a delusion: due to the multi-scaling nature of the stock market movements, even a movement of 20\% cannot be neglected;
Nassim Nicholas Taleb (cf.~\cite{taleb}) has compared such a delusion with the Russian Roulette 
with a revolver containing 500 bullet-holes: the fact that nothing has happened during the 100 first clicks on the trigger, does not make the game harmless: 
the 101$^{st}$ click can be lethal.

\section*{Acknowledgement}
The support of Estonian SF grant No. ETF6121 is acknowledged. We would also like to thank Prof. J\"uri Engelbrecht for fruitful discussions.

\appendix
\section{Appendix}
\label{appendix}
In order to test the equation (\ref{silence}), artificial data  with know statistical properties  we generated as follows.
To begin with, note that actual time-series are not needed; instead, it suffices to generate the sequence of the durations $\tau_i$ ($i=1,2,\ldots$) of the low-variability periods.
Indeed, our aim is to analyse the "silence-breaking" probability $p(n)$ for a fixed set of parameters $w$ and $\delta$ (without intention to repeat the test 
for the same time-series with different  parameter values).

So, the total number of low-variability intervals was fixed to be $L=10^6$ for Fig.~\ref{fig:surrogate-alphas} and $L=10^3$ for Fig.~\ref{fig:p-n-SPX-2stddev}. 
Further, for every index $i \in (0, L)$ the random power-law-distributed distributed number $\tau_i$ was obtained. 
The values of $\tau_i$ were interpreted as the lengths of low-variability periods. 
Finally, all the $L$ intervals were sorted according to their length and cumulative distribution function $R(n)$ was calculated.

\end{document}